\documentstyle[preprint,aps]{revtex}
\def\lesssim{\ ^< \!\!\!\!_{\sim}\ }
\def\gtrsim{\ ^> \!\!\!\!_{\sim}\ }
\draft
\begin{document}

\title{Effective Hamiltonian for Excitons with Spin Degrees of Freedom
}
\author{Jun-ichi Inoue\cite{email1}, 
Tobias Brandes\cite{newaddress}, 
and Akira Shimizu}
\address{Institute of Physics, University of Tokyo,
3--8--1 Komaba, Meguro-ku, Tokyo 153--8902, Japan
\\
and Core Research for Evolutional
Science and Technology, JST}
\date{\today}
\maketitle
\begin{abstract}
Starting from the conventional electron-hole Hamiltonian
${\cal H}_{eh}$, 
we derive an effective Hamiltonian $\tilde{\cal H}_{1s}$ 
for $1s$ excitons with spin degrees of freedom.
The Hamiltonian describes optical processes 
close to the exciton resonance for the case of weak excitation.  
We show that straightforward bosonization 
of ${\cal H}_{eh}$ does not give the correct form of 
$\tilde{\cal H}_{1s}$, which we obtain by a projection onto 
the subspace spanned by the $1s$ excitons.
The resulting relaxation and renormalization terms generate an
 interaction between excitons with opposite spin.  
Moreover, exciton-exciton repulsive interaction is greatly reduced by
the renormalization.  
The agreement of the present theory with the experiment
supports the 
validity of the description of a fermionic system by bosonic fields
in two dimensions.
\end{abstract}


\maketitle

Properties of a fermionic system can sometimes be described by bosonic
fields, by which theoretical analyses may be greatly simplified.  
The most successful example is the bosonization of one-dimensional ($d=1$)
conductors \cite{Stone}, where
success is due to the specific feature of the pair
spectrum \cite{Haldane}, i.e., the low-energy pair 
spectra consist of discrete 
branches in the energy versus momentum plane.
For $d \geq 2$ conductors, discrete branches overlap with 
continuous spectra, hence the bosonization is nontrivial
and still in progress \cite{Kopietz}.

For insulating solids, on the other hand,
discrete branches of excitons are separated from the continuum 
for any $d$.
From this point of view, 
it has been suggested that a useful bosonic 
theory may be constructed if one focuses on
exciton states, even for $d \geq 2$
\cite{Hanamura,Kuwata-Gonokami,Axt,Ivanov}.  
However, the validity of the bosonic description of excitons 
is nontrivial, because the binding energy (in, {\em e.g.}, GaAs) 
is comparable to other relevant energies \cite{He}.
Under an optical excitation, 
excitons and free electron-hole ({\em e-h}) pairs (continuous spectra) 
will be created.
As the excitation intensity (and thus the {\em e-h} density) is 
increased, 
the fermionic nature of the system becomes more important, and bosonization requires more
bosonic fields. 
Such a strong-excitation regime has been successfully analyzed 
without the use of bosonization \cite{Haug,Lindberg,Hu,Rappen,Schafer}.
In a weak-excitation regime, on the other hand, 
it is expected that the system is well described 
by a small number of bosonic fields.  
If this is the case, the bosonic theory will provide
a powerful theoretical tool as well as a transparent physical view,
as in the case of the $d=1$ conductors \cite{Stone}.

To demonstrate the effectiveness of the bosonized theory, 
optical experiments may be more
convenient than the electron transport experiments, because
one can easily produce and detect two or more light beams, 
and obtain rich information from responses to the multi beams.
Moreover, one can 
easily control the polarization of each individual light beam,
which gives more detailed information.
Recently, 
by controlling the polarizations of two light beams, 
Kuwata-Gonokami {\em et al.} demonstrated 
experimentally that the polariton-polariton scatterings 
in a quantum well (QW) in an optical micro cavity 
are well described by a phenomenological bosonic Hamiltonian
in the weak-excitation regime
\cite{Kuwata-Gonokami,lowestorder}.
Their experiment 
strongly indicated the validity of 
a bosonic description for $d=2$.  
However, 
no theoretical studies  were reported which derive their phenomenological Hamiltonian from a
microscopic fermionic theory.  

In this letter, we derive an effective Hamiltonian of excitons from the
conventional {\em e-h} Hamiltonian.  
In particular, 
we will show that the derivation of an effective bosonic theory 
is nontrivial and not 
straightforward because a direct bosonization does not give 
correct results.
Our calculation and the agreement with the experimental data
\cite{Kuwata-Gonokami} directly prove the validity and relevance of a
bosonic description for $d=2$.

{\it Model ---}
We start from a Hamiltonian ${\cal H}_{eh}$ for 
an interacting electron-hole system in a QW.
\begin{eqnarray}
{\cal H}_{eh}&=&\sum_{i}\int {\rm d}x
\hat{\psi}^{\dagger}_{i}(x)
\left(-\frac{\hbar^{2}\nabla^{2}}{2m_{i}}+E_{i}\right)
\hat{\psi}_{i}(x)
\nonumber\\
&&+
\sum_{i,i'}\frac{z_{i}z_{i'}}{2}
\int {\rm d}x {\rm d}x'
\hat{\psi}_{i}^{\dagger}(x)
\hat{\psi}_{i'}^{\dagger}(x')
V({\bf r}_{i}-{\bf r}^{\prime}_{i'})
\hat{\psi}_{i'}(x')
\hat{\psi}_{i}(x).
\label{Heh}
\end{eqnarray}
Here, $V({\bf r})$ denotes the Coulomb potential,
which behaves in a QW of width $L$ as $V({\bf r}) \approx e^2/ \epsilon r$ for 
$|{\bf r}| \gtrsim L$, where $\epsilon$ is the static dielectric constant, and 
$V({\bf r}) \approx$ constant for $|{\bf r}| \lesssim L$.
We simplify the calculation by taking 
the limit $L \to 0$ wherever the singularity at ${\bf r} = {\bf 0}$ 
is irrelevant.
In eq.\ (\ref{Heh}), $\hat{\psi}_{e(h)}(x)$ is the field operator of
an electron (hole), $\{i,i'\}=\{e,h\}$, $z_{e(h)}=1\,(-1)$, 
$x\equiv({\bf r}_{e(h)},J^{z}_{e(h)})$, and $\int {\rm
d}x\equiv\sum_{J^{z}_{i}}\int {\rm d}^{2}r_{i}$.  
The index $J^{z}_{e(h)}$ denotes 
the $z$-component of the total angular momentum (which is a good quantum
number), 
where we take the $z$-axis to be in the direction normal to the QW
layers.
The $J^{z}_{h}$ is defined as $-1$ times $J^{z}$
of a valence band electron.  
In a GaAs QW, {\em e.g.}, $J^{z}_{h}=\pm3/2$ 
for a heavy hole, and $J^{z}_{e}=\pm1/2$ \cite{Haug}.
A photon with $J^{z}_{ph}=+1(-1)$
creates an electron-hole pair with $J^{z}_{e}=-1/2\, (+1/2)$ for the
electron and 
with $J^{z}_{h}=+3/2\,(-3/2)$ for the hole. 
The exciton states are labelled by indices ${\bf q}$, $\nu$ and
$S$, where ${\bf q}$ is the momentum of the center-of-mass motion, 
$\nu$
denotes the set of quantum numbers for the relative motion 
($\nu = 1s, 2p_{+}, 2p_{-},\cdots$)\cite{Haug}, 
and $S$ denotes combinations of
$J^{z}_{e}$ and $J^{z}_{h}$.
Since $S$ is related to the total angular momentum, 
we call $S$ the ``spin'' index.

Since all states excited by photons are electrically neutral,
we can confine ourselves to the charge-neutral sector.  
We first transform ${\cal H}_{eh}$ to a Hamiltonian ${\cal
H}$ of interacting excitons.
We then make a projection onto 
the subspace spanned by 1s excitons 
to obtain an effective theory which is described by
a renormalized Hamiltonian $\tilde {\cal H}_{1s}$ of 1s excitons and a 
relaxation ${\tilde\Gamma}$.

{\it Transformation of ${\cal H}_{eh}$ into ${\cal H}$ ---}
We first rewrite ${\cal H}_{eh}$ as a function of exciton operators
\cite{Haug-Schmitt-Rink}, 
\begin{eqnarray}
b_{{\bf q}\nu S} &\equiv& \sum_{J^{z}_{e},J^{z}_{h}}
\int {\rm d}^2 r_{e}{\rm d}^2 r_{h}
\frac{1}{\sqrt{\Omega}}
\exp \left(
{i{\bf q} \cdot {m_e {\bf r}_{e}+ m_h {\bf r}_{h} \over M} }
 \right)
\nonumber\\
&& \times
\varphi_{\nu}({\bf r}_{e}-{\bf r}_{h})
\langle S|J^{z}_{e},J^{z}_{h}\rangle
\hat{\psi}_{e}({\bf r}_{e},J^{z}_{e})\hat{\psi}_{h}({\bf r}_{h},J^{z}_{h}).
\end{eqnarray}
Here, $\varphi_{\nu}({\bf r})$ is a wavefunction for the {\em e-h} relative motion, 
$\langle S|J^{z}_{e},J^{z}_{h}\rangle$ the Clebsch-Gordan (CG)
coefficient,
$\Omega$ the QW area, and $M \equiv m_{e}+m_{h}$.  
In the following, we assume $0 < m_e \ll m_h$.
To define $S$ uniquely, we choose  
$S=+,-,\alpha,$ and $\beta$,
for 
$|J^{z}_{e},J^{z}_{h}\rangle =$
$|-1/2,+3/2\rangle$,
$|+1/2,-3/2\rangle$, $|+1/2,+3/2\rangle$,
and $|-1/2,-3/2\rangle$, respectively, in a GaAs QW \cite{similar}.
An exciton with $S=\pm$ couples to circularly polarized light with  
$J^{z}_{ph} = \pm1$, 
whereas excitons with $S=\alpha$ or $\beta$ 
are dipole inactive. 
Then, assuming that the exciton density is low, 
we regard $b_{{\bf q}\nu S}$'s 
as genuine boson operators \cite{Haug-Schmitt-Rink} and
rewrite ${\cal H}_{eh}$ in the charge-neutral sector as 
${\cal H}_{eh}\to {\cal H}={\cal H}^{0}+{\cal H}^{int}$, where ${\cal
H}^{0}$ and ${\cal H}^{int}$ are the free and interaction parts of excitons,
respectively.  
Since the transferred momentum in the exciton scattering processes 
is fairly small, and of the order of the photon momentum,
the direct and the double fermionic exchange interactions are
negligible and the momentum dependence of the exchange interaction can be omitted.  
Hence, the general form of ${\cal H}^{int}$ is given by
\begin{eqnarray}
{\cal H}^{int}&=&\sum_{{\bf k}{\bf k}'{\bf q}\,\{S\}\{\nu\}}\frac{U_{\nu}}{2\Omega}
\sum_{J^{z}_{e}J^{z}_{h}J^{z\prime}_{e}J^{z\prime}_{h}}
 \langle S_{1}|J^{z}_{e}J^{z}_{h}\rangle
 \langle S_{2}|J^{z\prime}_{e}J^{z\prime}_{h}\rangle
 \langle S_{3}|J^{z\prime}_{e}J^{z}_{h}\rangle
 \langle S_{4}|J^{z}_{e}J^{z\prime}_{h}\rangle
\nonumber\\
&&\times
b^{\dagger}_{{\bf k}+{\bf q}\nu_{1} S_{1}}
b^{\dagger}_{{\bf k}'-{\bf q}\nu_{2} S_{2}}
b_{{\bf k}'\nu_{4} S_{4}}
b_{{\bf k}\nu_{3} S_{3}}.
\label{gint}
\end{eqnarray}
Using this formula, we express ${\cal H}$ as
\begin{equation}
{\cal H}
=
{\cal H}^{0}
+
{\cal H}^{\pm}_{1s} 
+ {\cal H}^{\prime}_{1s}
+{\cal H}_{others}, 
\label{Hx}\end{equation}
where ${\cal H}^{\pm}_{1s}$ and ${\cal H}^{\prime}_{1s}$ include the $\nu =
1s$ operators only, and ${\cal H}_{others}$ denotes the remaining terms.  
The ${\cal H}^{\pm}_{1s}$ consists
only of the operators with $S=\pm$, whereas 
${\cal H}^{\prime}_{1s}$ consists of terms
of $S=\alpha$ and $\beta$ operators, including
cross terms with $S=\pm$ operators:
\begin{eqnarray}
&{\cal H}^{\pm}_{1s}&
=
\frac{U}{2 \Omega}\sum_{S=\pm} \sum_{{\bf k}{\bf k}'{\bf q}}
b^{\dagger}_{{\bf k}+{\bf q}\,S}b^{\dagger}_{{\bf k}'-{\bf q}\,S}
b^{\phantom{\dagger}}_{{\bf k}'\,S}b^{\phantom{\dagger}}_{{\bf k}\,S},
\label{Hexpm}\\
{\cal H}^{\prime}_{1s}
&=&
{U \over \Omega} \sum_{{\bf k}{\bf k}'{\bf q}} 
\Biggl[
\sum_{S = \alpha, \beta} 
\left(
\frac{1}{2} b^{\dagger}_{{\bf k}+{\bf q}\,S}b^{\dagger}_{{\bf k}'-{\bf q}\,S}
b^{\phantom{\dagger}}_{{\bf k}'\,S}b^{\phantom{\dagger}}_{{\bf k}\,S}
\right.
+ 
b^{\dagger}_{{\bf k}+{\bf q}\,+}b^{\dagger}_{{\bf k}'-{\bf q}\,S}
b^{\phantom{\dagger}}_{{\bf k}'\,S}b^{\phantom{\dagger}}_{{\bf k}\,+}
\nonumber\\
&&+
 \left.
b^{\dagger}_{{\bf k}+{\bf q}\,-}b^{\dagger}_{{\bf k}'-{\bf q}\,S}
b^{\phantom{\dagger}}_{{\bf k}'\,S}b^{\phantom{\dagger}}_{{\bf k}\,-}
\right)
+
\left(
b^{\dagger}_{{\bf k}+{\bf q}\,+}b^{\dagger}_{{\bf k}'-{\bf q}\,-}
b^{\phantom{\dagger}}_{{\bf k}'\,\alpha}b^{\phantom{\dagger}}_{{\bf k}\,\beta}
+ {\rm H.c.}
\right)
\Biggr],
\label{int-ign}
\end{eqnarray}
where we have written $b_{{\bf q} 1s S} \equiv b_{{\bf q} S}$,
and the effective 
interaction strength $U\equiv U_{\nu=1s}$ is expressed as
\begin{equation}
U=
2 \sqrt{\Omega}
\sum_{{\bf p},{\bf p}'}
\tilde{V}({\bf p}-{\bf p}')
\Bigl[
|{\tilde\varphi}_{1s}({\bf p})|^{2}
{\tilde\varphi}^{*}_{1s}({\bf p})
{\tilde\varphi}_{1s}({\bf p}')
-
|{\tilde\varphi}_{1s}({\bf p})|^{2}
|{\tilde\varphi}_{1s}({\bf p}')|^{2}
\Bigr].
\label{U}
\end{equation}
Here, $\tilde{V}({\bf p})$ and $\tilde{\varphi}_{\nu}({\bf p})$ are 
the Fourier transforms of
$V({\bf r})$ and $\varphi_{\nu}({\bf r})$, respectively.
For $d=2$, and in the limit of $L \to 0$, 
eq.\ (\ref{U}) is evaluated as
$
U= 1.52 a^{2}_{0} E^{b}_{x}
$ \cite{Schmitt-Rink2}, 
where $a_{0} \equiv \epsilon \hbar^2 M/e^2 m_e m_h$ and 
$E^{b}_{x} \equiv 2 e^2 / \epsilon a_0$.

It is tempting to take
$
{\cal H}^{\pm}_{1s}+{\cal H}^{\prime}_{1s}
$ 
as an effective interaction Hamiltonian ${\cal H}^{eff}_{1s}$ 
for $1s$ excitons, 
by simply dropping 
${\cal H}_{others}$. 
However, as we will show below, this is not an appropriate approximation of
${\cal H}^{eff}_{1s}$, because 
important terms which describe the interactions between excitons with
$S=+$ and with $S=-$ are absent in ${\cal H}^{\pm}_{1s}+{\cal
H}^{\prime}_{1s}$ \cite{Ivanov2}.

{\it Projection ---}
The evolution of the density operator $\rho$ 
in the charge neutral sector 
obeys the von Neumann equation,
$\partial \rho/\partial t=(1/i\hbar)[{\cal H}, \rho]$.
To derive the correct 
${\cal H}^{eff}_{1s}$, 
we make a projection onto 
the subspace spanned by the 1s excitons.
We then obtain the equation for 
the reduced density operator $\tilde\rho_{1s}$, as
$
\partial \tilde\rho_{1s}/\partial t
=(1/i\hbar)[{\tilde{\cal H}}_{1s}, \tilde\rho_{1s}]
+ \tilde\Gamma \tilde\rho_{1s}
$.
Here, ${\tilde{\cal H}}_{1s}$ describes the unitary evolution
of $\tilde\rho_{1s}$, and $\tilde\Gamma$ the relaxation operator.
We can calculate optical responses using ${\tilde{\cal H}}_{1s}$, neglecting
$\tilde\Gamma$, when (a) the photon energy is close to the 
energy of the $1s$ excitons,
(b) the photoexcitation is weak, 
and (c) relaxation processes do not
play an important role during the photon scattering processes.
The reason for (a) is that $2p$ and higher excitons have been projected out.  
The condition (b) is due to the fact that the deviation from the Bose
statistics of operators $b_{{\bf q}\nu S}$ and $b^{\dagger}_{{\bf q}\nu S}$ 
becomes non-negligible when the
excitation is strong \cite{Hawton}. 
In the experiment described in ref.\  \cite{Kuwata-Gonokami}, 
the three conditions have been satisfied.  
To satisfy condition (c), both the resonance of $1s$ excitons and
the strong coupling to the
radiation field in a high Q micro cavity are crucial \cite{Kuwata-Gonokami1}.  
Otherwise, we must take account of ${\tilde \Gamma}$ in the calculations
of optical responses.

The renormalization procedure yields \cite{unpublished}
\begin{equation}
{\tilde{\cal H}}_{1s}
=
{\tilde{\cal H}}^{0}_{1s}
+
{\tilde{\cal H}}^\pm_{1s}+{\tilde{\cal H}}'_{1s},
\label{H1s}\end{equation}
where
${\tilde{\cal H}}^{0}_{1s}$ is the Hamiltonian of free $1s$ excitons, 
and
${\tilde{\cal H}}^{\pm}_{1s}$ and ${\tilde{\cal H}}'_{1s}$
include the $\nu = 1s$ operators only.  
The ${\tilde{\cal H}}^{\pm}_{1s}$ consists
only of operators with $S=\pm$, whereas
${\tilde{\cal H}}^{\prime}_{1s}$ consists of terms
of $S=\alpha$ and $\beta$ operators, including
cross terms with $S=\pm$ operators. 
Since 
${\tilde{\cal H}}'_{1s}$ includes 
dipole inactive $1s$ 
excitons ($S=\alpha$, $\beta$),
it does not contribute to the optical response 
in its lowest order \cite{lowestorder,inactive}.  
The most important terms are therefore included in 
${\tilde{\cal H}}^\pm_{1s}$, which 
is evaluated, to the second order in the 
exciton-exciton interactions, as
\begin{equation}
{\tilde{\cal H}}^\pm_{1s}=\frac{U-U'}{2 \Omega}
\sum_{S=\pm}\sum_{{\bf k}{\bf k}'{\bf q}}
b^{\dagger}_{{\bf k}+{\bf q}\,S}b^{\dagger}_{{\bf k}'-{\bf q}\,S}
b^{\phantom{\dagger}}_{{\bf k}'\,S}b^{\phantom{\dagger}}_{{\bf k}\,S}
- {U' \over \Omega}
\sum_{{\bf k}{\bf k}'{\bf q}}
b^{\dagger}_{{\bf k}+{\bf q}\,+}b^{\dagger}_{{\bf k}'-{\bf q}\,-}
b^{\phantom{\dagger}}_{{\bf k}'\,-}b^{\phantom{\dagger}}_{{\bf k}\,+},
\label{int-1s}
\end{equation}
where $U'$ is a positive constant
which arises from the renormalization of 
higher exciton states
$(\nu=2p_{+},\,2p_{-},\,\cdots)$:
\begin{eqnarray}
U'
&=&
\Omega
\sum_{{\bf K},\nu\neq 1s}
\frac{1}{2\left(E_{\nu}+{\bf K}^2/2M\right)-2E_{1s}}
\left|
\sum_{{\bf p},{\bf p}'}
\tilde{V}({\bf p}-{\bf p}'+{\bf K})
\Bigl[
-\tilde{\varphi}^{*}_{1s}({\bf p})
 \tilde{\varphi}^{*}_{1s}({\bf p}')
 \tilde{\varphi}_{\nu}({\bf p})
 \tilde{\varphi}_{\nu}({\bf p}')
\right.
\nonumber\\
&&\hspace{-3mm}\phantom{\sum_{{\bf p}}}\phantom{|}
+2\tilde{\varphi}^{*}_{1s}({\bf p})\tilde{\varphi}^{*}_{1s}({\bf p}-{\bf K})
\tilde{\varphi}_{\nu}({\bf p})\tilde{\varphi}_{\nu}({\bf p}')
\left.
-\tilde{\varphi}^{*}_{1s}({\bf p})\tilde{\varphi}^{*}_{1s}({\bf p}')\tilde{\varphi}_{\nu}({\bf p}-{\bf K})
\tilde{\varphi}_{\nu}({\bf p}'+{\bf K})\Bigr]\phantom{\sum_{{\bf p}}}\hspace{-5mm}\right|^{2}. 
\label{U'}
\end{eqnarray}
Comparing the right-hand side of eq.\ (\ref{int-1s}) 
with that of eq.\ (\ref{Hexpm}), 
we find that 
the coefficient of the first term is renormalized as 
$U \to U-U'$, and that a  second term is generated which leads to an
interaction between the $S=+$ and $-$ excitons. 
That is, 
the renormalization of 
higher exciton states results in 
the renormalized Hamiltonian ${\tilde{\cal H}}^\pm_{1s}$,
which differs, both quantitatively and qualitatively, 
 from the bare Hamiltonian ${\cal H}^{\pm}_{1s}$.

We argue that the correct form of the 
effective Hamiltonian for $1s$ excitons is 
the renormalized one, i.e.,   
${\cal H}^{eff}_{1s} = {\tilde{\cal H}}^\pm_{1s} + {\tilde{\cal
H}}'_{1s}$.   
In fact,  
the interaction between the $S=+$ and $-$ excitons in ${\tilde{\cal H}}_{1s}^{\pm}$, 
which is absent in ${\cal H}_{1s}$, 
has been clearly observed experimentally in refs.\ 
\cite{Kuwata-Gonokami} and \cite{Bott}.
Kuwata-Gonokami {\em et al.} \cite{Kuwata-Gonokami} 
expressed this interaction as 
an interaction term (whose coupling constant is $W$) in the phenomenological Hamiltonian, which also has
the interaction term (whose coupling constant is $R$) of excitons with
parallel spins.  
The phenomenological Hamiltonian 
has the same form as ${\tilde{\cal H}}_{1s}^{\pm}$, 
the dipole active part of $\tilde{\cal H}_{1s}$.
This is quite reasonable because the other part ${\tilde{\cal H}}'_{1s}$, 
which is dipole inactive, 
should be invisible in low-order optical experiments \cite{inactive}.
We can, therefore, identify the parameters $R$ and $W$ of the
phenomenological Hamiltonian
\cite{Kuwata-Gonokami} as
\begin{eqnarray}
R &=& (1.52 a^{2}_{0} E^{b}_{x}-U')/(2\Omega),
\label{R}\\
W &=& -U'/\Omega.
\label{W}\end{eqnarray}
The value of $U'$, as given by eq.\ (\ref{U'}),  
depends on the material parameters such as 
$M$ and $\epsilon$, and hence it is different for different 
materials.
It also depends on the QW parameter $L$.
Moreover, when imperfections in the QW are non-negligible, 
the formula for $U'$ should be modified accordingly.
Therefore, even for the same material, 
the values of $R$ and $W$ could vary slightly from sample to sample,
which seems to be consistent with recent experimental results \cite{Kuwata-Gonokami1}.
Note, however, that the {\em existence} of both terms of ${\tilde{\cal H}
}^{\pm}_{1s}$ is independent of such details.  

We here estimate the typical value of $U'$ as follows.
The ${\bf K}$-summation in eq.\ (\ref{U'}) is cutoff 
for $K \gtrsim C_L / L$ (through the cutoff of ${\tilde V}$) 
and/or for $K \gtrsim C_{a_0} / a_0$ (through $\tilde \varphi_\nu$),
where $C_L$ and $C_{a_0}$ are cutoff parameters of the order of unity.
For the case of the QW sample of 
ref.\ \cite{Kuwata-Gonokami}, 
$L \approx a_0$, hence we may cutoff 
the ${\bf K}$-summation 
for $K \gtrsim C / a_0$, where $C$ is of the order of unity.
For the $\nu$ summation, 
we may consider
$\nu=2p_{\pm}$ states only, because higher exciton states 
give much smaller overlap integrals.
These approximations yield
$
 U'\approx
16.5 C^2 a^{2}_{0} E^{b}_{x}.  
$
On the other hand, 
ref.\ \cite{Kuwata-Gonokami} reported the ratio
$R:W$ as $1:-15$.
From eqs.\ (\ref{R}) and (\ref{W}), 
we find that this ratio is reproduced by the present theory
when the cutoff parameter $C \sim 0.3$, 
which is consistent with the requirement that $C$ is of the order of unity.
Considering that the values of $R$ and $W$ vary slightly from sample
to sample \cite{Kuwata-Gonokami1}, the agreement seems satisfactory.
Note that such a small value of $R$ is due to the renormalization of $U\to U-U'$.  
Once the agreement of ${\tilde{\cal H}}_{1s}^{\pm}$ 
with the phenomenological Hamiltonian is 
thus established, 
the agreement with the experiment follows.
That is, 
lowest-order perturbational calculations
for the polariton-polariton scattering amplitudes  
agree with the experiment \cite{Kuwata-Gonokami,lowestorder}.

{\it Discussions and remarks ---}
It was conjectured \cite{Kuwata-Gonokami} that 
a ``biexciton effect'' would be 
the origin of the ``{\em W} term'', 
the interaction between $S = +$ and $-$ excitons.
However, this argument is misleading.
The biexciton state is formed 
essentially from the mixing of two $1s$ states having different centers.
For examples, in the case of a hydrogen molecule,
the mixing yields the bonding and antibonding 
states,
$
(1/\sqrt{2})
(c^{\dagger}_{1\uparrow}c^{\dagger}_{2\downarrow}
\pm
c^{\dagger}_{1\downarrow}c^{\dagger}_{2\uparrow})
h^{\dagger}_{1\sigma}h^{\dagger}_{2\sigma'}|0\rangle
$.
Here, $c^\dagger_{1 (2)}$ creates an electron in the $1s$ state
located at nucleus 1(2), 
and  $h^\dagger_{1(2)}$ creates the nucleus.
In the case of excitons with $J^{z}_{e}=\pm1/2$ and $J_h^{z} = \pm 3/2$
\cite{similar},
the corresponding states are
$
(1/\sqrt{2})[
b^{\dagger}_{+}b^{\dagger}_{-}
\pm
b^{\dagger}_{\alpha}b^{\dagger}_{\beta}
]|0\rangle
$,
where we have not shown the ${\bf k}$-dependence in order to focus on 
the $S$-dependence.
The bonding state ($-$ sign for a positive coupling constant) has a
lower energy and is called a biexciton. 
This energy splitting 
between the bonding and antibonding states
is induced by 
the interaction of the form of 
$b_+^\dagger b_-^\dagger b_\alpha b_\beta + H.c.$, 
which is included in 
$\tilde{\cal H}'_{1s}$  (or, before the renormalization, in 
${\cal H}^{\prime}_{1s}$ of eq.\ (\ref{int-ign})) \cite{biexciton}.
On the other hand, 
the $W$ term lowers the energies of {\em both} states
{\em by the same amount}, hence 
does not play a central role
in the formation of the biexciton state.
The most important effect of the $W$ term is 
to lower the energy of 
$b^{\dagger}_{+}b^{\dagger}_{-}|0\rangle$, relative to those of 
$b^{\dagger}_{+}b^{\dagger}_{+}|0\rangle$ and
$b^{\dagger}_{-}b^{\dagger}_{-}|0\rangle$, 
and this effect was detected experimentally \cite{Kuwata-Gonokami}.  
In short, in the framework of the present theory, 
$\tilde{\cal H}'_{1s}$ lowers the energy of the bonding (biexciton)
state relative to that of the antibonding state, and thus is crucial for
the formation of the biexciton state,  whereas the ${\em W}$
term lowers the energy of both bonding and antibonding states.

Note that 
${\tilde{\cal H}}_{1s}$ is not positive definite to the fourth
 order in the exciton operators.
The stability of the system should be preserved by higher order terms.
In general situations, 
properties of a system described by such a Hamiltonian 
should not be analyzed by a perturbation theory based on 
the vacuum of the free part. 
Nevertheless, we can use such a perturbation theory 
in our case, because our 
exciton theory has the built-in constraint that the ground state 
is the state with no excitons, i.e., the vacuum of ${\tilde{\cal H}}^{0}_{1s}$.
The effective Hamiltonian ${\tilde{\cal H}}_{1s}$ together with this constraint 
constitutes a consistent theory, which justifies 
the low-order perturbation theory based on the given vacuum, if
the optical excitation is sufficiently weak.

We have used a low-order perturbation theory 
to derive $\tilde{\cal H}_{1s}$.
However, this does not imply a total neglect of
higher order terms, 
because we have calculated a Hamiltonian rather than observables.
In fact, 
a systematic summing up of higher order terms is already incorporated in our theory 
if one calculates 
higher order scattering amplitudes, {\em e.g.}, by writing the Bethe-Salpeter
equation and using ${\tilde{\cal H}}_{1s}$.

Finally, we discuss 
the relation between the fermionic theories 
\cite{Haug,Lindberg,Hu,Rappen,Schafer} and our bosonic theory.  
The Hartree-Fock (HF) factorization treatment
of the semiconductor Bloch equations \cite{Lindberg}
can not produce the interaction 
between the $S=+$ and $-$ excitons.
The HF theory, therefore, corresponds to
${\cal H}^{\pm}_{1s}$, eq.\ (\ref{Hexpm}).
It was argued in refs.\ \cite{Hu,Rappen,Schafer} that
the interactions of an exciton with higher states 
(including free carriers) 
are important, and that
the interactions result in
the energy shift, the
excitation-induced dephasing (EID), and the ``biexcitonic correlations''.
In the bosonic theory in the form of 
eq.\ (\ref{Hx}), these effects are included in 
${\cal H}^{\prime}_{1s}$
and 
${\cal H}_{others}$.
After the projection is made,
the relation is roughly as follows.
The renormalized Hamiltonian
${\tilde{\cal H}}^{\pm}_{1s}$, eq.\ (\ref{int-1s}),
would include
the HF term and a part of the ``biexcitonic correlation.''
The EID may be described by both
$\tilde\Gamma$ and ${\tilde{\cal H}}'_{1s}$.
Another part of the ``biexcitonic correlation'' would also be included 
in ${\tilde{\cal H}}'_{1s}$.
The present theory thus helps to bridge the gap between
the bosonic theories \cite{Hanamura,Kuwata-Gonokami,Axt,Ivanov} 
and the fermionic theories \cite{Haug,Lindberg,Hu,Rappen,Schafer}
of {\em e-h} systems.
However, more detailed comparisons will be a subject of future studies.

{\it Summary and Conclusions ---}
In this letter, starting
from the conventional electron-hole Hamiltonian (eq.\ (1)), we have
derived the effective Hamiltonian $\tilde{{\cal
H}}_{1s}$ for the $1s$ excitons with the spin degrees of freedom using the
bosonic exciton operators in two dimensions (eq.\ (2)).  
It is found that the renormalization associated with
the projection onto the $1s$ exciton space is
crucial, which leads to the generation of the
attractive interaction between excitons with
opposite spins (eq.\ (12)), and to the large reduction of the repulsive
interaction between excitons with parallel spins (eq.\ (11)).  
Such a drastic modification of the interactions was absent in the previous theory
without the renormalization procedure.  
The present theory is valid for systems that satisfy the following
conditions:
(i) excitation is weak, (ii) the $1s$ excitons play a crucial role, and
(iii) the exciton
relaxation process is less important due to, for example, the micro cavity
of a high Q-value.  
This effective Hamiltonian provides the
microscopic foundation of the phenomenology proposed in ref.\ \cite{Kuwata-Gonokami}.  
The agreement of the present theory with the experiment supports the 
validity of the description of a fermionic system by bosonic fields in two dimensions.

Helpful discussions with Professor Kuwata-Gonokami and
Dr.\ Suzuura are acknowledged.


\begin{thebibliography}{99}
\bibitem{He}
This should be contrasted with liquid Helium 4:
there is no doubt that an Helium 4 atom  
can be described by a
boson operator when the superfluidity is studied, 
because 
its nuclear binding energy is much larger
than typical energy scales of superfluid experiments.
\bibitem{inactive}However, ${\tilde{\cal H}}^{\prime}_{1s}$, of course, becomes important 
when one measures optical processes in which 
higher-order terms are involved.
\bibitem{biexciton}
A typical bound state in an interacting boson model is the eigenstate
$|\phi\rangle=(a^{\dagger}b^{\dagger}-{\rm sign}(g) c^{\dagger}d^{\dagger})|0\rangle$ of
${\cal H}=\omega(a^{\dagger}a+b^{\dagger}b+c^{\dagger}c+d^{\dagger}d)+
g(a^{\dagger}b^{\dagger}cd+d^{\dagger}c^{\dagger}ba)$.
The eigenenergy is $2\omega-|g|$.
Note that the existence of the bound state is independent of the sign of
	   the interaction $g$.
\end{thebibliography}
\end{document}